\documentclass[preprint,showpacs, superscriptaddress, preprintnumbers,amsmath,amssymb]{revtex4}

\usepackage{booktabs}
\usepackage{color,graphicx}
\usepackage{amsmath}
\usepackage{dcolumn}
\usepackage{bm}                  
\usepackage{soul}
\usepackage{enumerate}
\usepackage{tabularx}
\usepackage{multirow}
\usepackage{threeparttable}

\newcommand{\bk}{\mbox{\boldmath$k$}}
\newcommand{\bsig}{\mbox{\boldmath$\sigma$}}
\newcommand{\half}{\frac{1}{2}}
\newcommand{\bnab}{\mbox{\boldmath$\nabla$}}
\newcommand{\bJ}{\mbox{\boldmath$J$}}
\def\be{\begin{equation}}
\def\ee{\end{equation}}
\def\bea{\begin{eqnarray}}
\def\eea{\end{eqnarray}}
\def\bse{\begin{subequations}}
\def\ese{\end{subequations}}
\def\br{{\bf r}}

\begin{document}

\title{Hyperon halo structure of C and B isotopes}

\author{Ying Zhang}
\email{yzhangjcnp@tju.edu.cn}
\affiliation{Department of Physics, School of Science, Tianjin University,
	Tianjin, 300354, China}
\affiliation{RIKEN Nishina Center, Wako, Saitama 351-0198, Japan}

\author{Hiroyuki Sagawa}
\email{sagawa@ribf.riken.jp}
\affiliation{RIKEN Nishina Center, Wako, Saitama 351-0198, Japan}
\affiliation{Center for Mathematics and Physics, the University of Aizu,
Aizu-Wakamatsu, Fukushima 965-8580, Japan}

\author{Emiko Hiyama}
\email{hiyama@riken.jp}
\affiliation{Department of Physics, Tohoku University, 980-8578, Japan}
\affiliation{RIKEN Nishina Center, Wako, Saitama 351-0198, Japan}

\begin{abstract}
	We study the $\Lambda$ hypernuclei of C and B isotopes by Hartree-Fock model  with Skyrme-type nucleon-nucleon and $\Lambda$-nucleon interactions.  The calculated $\Lambda$ binding energies agree well with the available experiment data.   
	We found halo structure in the $\Lambda$ $1p$-state with extended wave function beyond nuclear surface in the light C and B isotopes.  We also found the enhanced electric dipole transition between $\Lambda$
	$1p$- and $1s$-states, which could be the evidence for this hyperon halo structure.
\end{abstract}

\date{\today}

\pacs{
21.80.+a,  
21.60.Jz, 
21.10.Gv. 
}

\maketitle

\section{Introduction}
Since the halo structure of $^{11}$Li was observed in 1985 \cite{Tanihata},  the halo phenomena have been studied intensively from both experimental and
theoretical sides  \cite{Jensen2004RMP,Jonson2004PR,Hagino,Meng2015JPG} in the nuclei near and beyond  the neutron and also proton drip lines.  The halo nuclei are characterized by its extended density profile far beyond the nuclear surface region.  Very much enhanced electric dipole 
transitions have been also observed in several halo nuclei as an unique phenomenon associated with the extended halo wave function \cite{Nakamura}.
As a theoretical model, for lighter nuclei such as $^6$He and $^{11}$Li,  the framework of
core$+n+n$ three-body model has been adopted often to describe the so called 
 "Borromean system",  in which one-nucleon+core system has never been bound, but only two-nucleon+core system makes a bound nucleus \cite{Fadeev,Bertsch}.
For $sd$-shell neutron-rich nuclei such as Ne isotope,
some halo states have been found~\cite{Nakamura2009PRL}. In addition, deformed structure
with larger $\beta_2$ has been observed in these systems  \cite{Takechi2012PLB}.
In the nuclei so far discussed, one or two nucleons will contribute to create the halo structure.
When one  goes to heavier nuclei, for instance, in neutron-rich Ca and Zr isotopes, theoretically in Refs.~\cite{Meng1998PRL,Meng2002PRC,Zhang2003SCG,Terasaki2006PRC,Grasso2006PRC,ZhangY2012PRC},
 giant halo nucleus is predicted,  in which several neutrons contribute to make halo nuclei.

Let us consider hypernuclei consisting  nuclei and a hyperon,
especially a $\Lambda$ particle.  So far, there have been many investigations on the effect of the hyperon in the neutron-rich hypernuclei~\cite{Vretenar1998PRC,Lu2003EPJA,Zhou2008PRC,Umeya2009 PRC,Gal2013PLB,Wirth2018PLB}, and even the explorations of the hyperon halo or hyperon drip line~\cite{Lu2002CPL,Khan2015PRC}.
Some authors pointed out that there were possibility to have 
halo states in lighter systems \cite{Miyagawa1995PRC,Hiyama1996PRC}:
In $^3_{\Lambda}$H, the observed binding energy is $0.13$ MeV 
with respect to deuteron+$\Lambda$ threshold,
which is a very weakly bound state and then this system has  
a $\Lambda$ halo structure with respect to deuteron \cite{Miyagawa1995PRC}.
One of the present authors (E. H.) pointed out that
  neutron or proton densities in the ground state of
$^6_{\Lambda}$He, 
excited states of $^7_{\Lambda}$He and
$^7_{\Lambda}$Li with isospin $T=1$ have been enhanced with the framework of
$^5_{\Lambda}{\rm He}+N+N$ three-body model \cite{Hiyama1996PRC}.  
Thus the study of halo structure in $\Lambda$ hypernuclei has been focused
on lighter hypernuclei with $A\leq 7$. 
In this paper, we focus on the possibility to have a  halo structure in heavier $\Lambda$ hypernuclei such as Boron or Carbon isotopes with $A\geq 8$.
Especially, in $^{13}_{\Lambda}$C, we have observed data of the ground state,
$1/2^+_1$, and either $3/2^+_1$ or $5/2^+_1$ positive-parity excited state,  $3/2^-$ and $1/2^-$ negative-parity excited states.  The 
dominate component of the two negative-parity states is $^{12}{\rm C}\otimes \Lambda (1p)$ configuration.
They are important to
extract the information on $\Lambda N$ spin-orbit force: 
they measured the spin-orbit splitting energy of $1/2^-$-$3/2^-$ to be 0.152 MeV
\cite{Ajimura2001PRL,Kohri-2002}.
Furthermore these states are weakly bound by about 1 MeV with respect to
$^{12}{\rm C}+ \Lambda$ threshold.
This means that we have a chance to find 
$\Lambda$ halo structure in C isotopes.
Therefore, in this paper, we focus on these possible $\Lambda$ halo states.
In addition, 
experimentally, a long isotope chain from $^{8}$C to $^{22}$C was observed.
Considering this situation, 
we study the ground states and the excited states (C $\otimes \Lambda(1p)$)
of C hypernuclei  systematically
with Hartree-Fock model using Skyrme-type nucleon-nucleon ($NN$) and $\Lambda$-nucleon ($\Lambda N$)
interactions, discuss on the halo structure of hypernuclei and
the possibility to observe these halo structures
by calculating the reduced transition probability $B(E1)$ from the $\Lambda(1p)$ state to the ground state $\Lambda(1s)$.


For this calculation, we use the Skyrme-Hartree-Fock model~\cite{Vautherin1972PRC}, which is commonly adopted for the description of the gross properties of the  nuclei in a broad  region of mass table.  The original Skyrme model has  no strangeness degree of freedom.  In 1981, Rayet introduced the Skyrme-type $\Lambda N$ interaction to describe the hypernuclei within the Skyrme model~\cite{Rayet1981NPA}.  Since then,  many  Skyrme-type $\Lambda N$ interactions were proposed based on  realistic hyperon-nucleon interactions,   stimulated by  many  hypernuclear data~\cite{Yamamoto1988PTP,Millener1988PRC,Fernandez1989ZPA,Lanskoy1997PRC,Cugnon2000PRC,Vidana2001PRC,Guleria2012NPA,Schulze2013PRC}.  With these interactions,  the hypernuclear structures have been investigated extensively~\cite{Zhou2007PRC,Zhou2008PRC,Win2011PRC,Li2013PRC}.   But most of these investigations did not include the $\Lambda N$ spin-orbit interaction, since it was expected to be rather small.  In this paper, we will adopt the Skyrme-type $\Lambda N$ interaction~\cite{Lanskoy1997PRC} obtained by the $G-$matrix calculation from the one-boson-exchange potential with a reduced $\Lambda N$ spin-orbit coupling  strength which can reproduce the spin-orbit splitting of the $1p$ states in $^{13}_{\Lambda}$C~\cite{Ajimura2001PRL}.  
The method is also applied to the neighboring Boron isotopes to discuss the $p$-wave $\Lambda$ hyperon halo structure there.
These studies are performed for the first time with this framework.
 
Organization of the present paper is as follows:
In Section II, the Method is explained.
The results are discussed in Sec. III and finally we summarize in Sec. IV.

\section{Theoretical Framework}
Hypernuclei of C and B isotopes are studied by using HF model with Skyrme-type $NN$ and $\Lambda N$ interactions. 
  The model is extended  to describe systematically  from light to heavy hypernuclei including the hyperon degree of freedom.
In the Skyrme model, the two-body  $NN$ interaction~\cite{Bender2003RMP} reads, 
\begin{eqnarray}
v_{NN}(\br_1-\br_2)  &=& t_0\left( 1+x_0P_{\sigma}\right)\delta(\br_1-\br_2)
+\half t_1\left(1+x_1 P_{\sigma}\right)\left[ \bk'^2\delta(\br_1-\br_2)+\delta(\br_1-\br_2)\bk^2\right] \nonumber \\
&&
+t_2\left(1+x_2 P_{\sigma}\right)\bk'\cdot\delta(\br_1-\br_2)\bk+iW_0(\bsig_1+\bsig_2) \cdot \bk' 
\delta(\br_1-\br_2)\times \bk,
\label{eq:Skyrme-NN}
\end{eqnarray}
where $\bk=(\overrightarrow{\bnab}_1-\overrightarrow{\bnab}_2)/2i$ is the relative momentum operator
acting on the wave functions on the right and $\bk'=-(\overleftarrow{\bnab}_1-\overleftarrow{\bnab}_2)/2i$ acting on the left, $P_{\sigma}= (1+\bsig_1\cdot \bsig_2)/2$ is the spin-exchange operator.  
The effective density-dependent  $NN$ interaction is also introduced as 
\begin{equation}  \label{V-den}
  v_{den-NN}(\br_1,\br_2,\br_3) = \frac{1}{6}t_3\left(1+x_3P_{\sigma}\right)\delta(\br_1-\br_2)\rho^{\alpha}\left(\frac{\br_1+\br_2}{2}\right),
\end{equation}
where $\alpha$ is the power of density dependence.  The Skyrme-type three-body force is equivalent to the interaction \eqref{V-den} with choices of $x_3=1$ and $\alpha=1$ for HF calculations. 

The Skyrme-like two-body $\Lambda N$ interaction is taken as~\cite{Lanskoy1997PRC}
\begin{eqnarray}
v_{\Lambda N}(\br_{\Lambda}-\br_N) &=& t^{\Lambda}_0(1+x^{\Lambda}_0P_{\sigma})\delta(\br_{\Lambda}-\br_{N})+\half t^{\Lambda}_1\left[ \bk'^2\delta(\br_{\Lambda}-\br_{N}) + \delta(\br_{\Lambda}-\br_{N})\bk^2 \right] \nonumber \\
&+& t^{\Lambda}_2 \bk'\delta(\br_{\Lambda}-\br_{N}) \cdot \bk + iW^{\Lambda}_0 \bk'\delta(\br_{\Lambda}-\br_{N}) \cdot (\bsig_N+\bsig_{\Lambda}) \times \bk \label{eq:Skyrme-Lambda-N}
\end{eqnarray}
with an effective density-dependent $\Lambda N$ force
\begin{equation}
v_{den-\Lambda N} (\br_{\Lambda}, \br_{N}, \rho) = 
\frac{3}{8} t^{\Lambda}_3(1+x^{\Lambda}_3P_{\sigma})\delta(\br_{\Lambda}-\br_N)\rho^{\gamma}\left( \frac{\br_{\Lambda}+\br_N}{2}\right), 
\end{equation}
where $\gamma$ is the power of density dependence. 

The total energy functional can be separated into two parts,
\begin{eqnarray}
E = \int d\br (\mathcal{H}_N + \mathcal{H}_{\Lambda}),
\end{eqnarray}
where $\mathcal{H}_N$ is the hamiltonian density only related with the nucleons, 
and $\mathcal{H}_{\Lambda}$ is the one with $\Lambda$ hyperon degree of freedom.   
 The nucleon hamiltonian density  $\mathcal{H}_N$ can be written as
 \begin{eqnarray}
 \mathcal{H}_N 
  &=& \frac{\hbar^2}{2m_N}\tau_N 
 + \frac{1}{2} t_0 \left( 1+\frac{1}{2}x_0\right) \rho^2_N
 - \frac{1}{2} t_0\left(x_0 + \frac{1}{2}\right)(\rho_n^2+\rho_p^2)  \nonumber \\
 && + \frac{1}{4}\left[t_1\left(1+\frac{1}{2}x_1\right) + t_2\left(1+\frac{1}{2}x_2\right)\right] \rho_N\tau_N
 + \frac{1}{4}\left[-t_1\left(\frac{1}{2}+x_1\right) + t_2\left(\frac{1}{2}+x_2\right)\right]\left( \rho_n\tau_n+\rho_p\tau_p\right)  \nonumber \\
 && + \frac{1}{16} \left[3t_1\left(1+\frac{1}{2}x_1\right)-t_2\left(1+\frac{1}{2}x_2\right)\right](\bnab\rho_N)^2   \nonumber \\
 && -  \frac{1}{16} \left[3t_1\left(\frac{1}{2}+x_1\right)+t_2\left(\frac{1}{2}+x_2\right)\right] \left[ (\bnab\rho_n)^2+(\bnab\rho_p)^2\right]  \nonumber \\
 && 
 + \frac{1}{16}\left[\left(  t_1-t_2\right)\left( \bJ_n^2+\bJ_p^2\right)
 -\left(t_1x_1+t_2x_2\right)  \bJ^2_N \right]  \nonumber \\
 && 
 + \frac{1}{12}t_3\left(1+\frac{1}{2}x_3 \right)\rho^{\alpha+2}_N
 - \frac{1}{12}t_3\left(\frac{1}{2}+x_3 \right)\rho^{\alpha}_N\left(\rho_n^2+\rho_p^2\right) \nonumber \\
 &&
 + \frac{1}{2}W_0\left(\bnab\rho_N\cdot\bJ_N+\bnab\rho_n\cdot\bJ_n
 +\bnab\rho_p\cdot\bJ_p\right) + \mathcal{H}_{\rm coul.}.    
 \label{eq:total-H-N}
 \end{eqnarray} 
In Eq. \eqref{eq:total-H-N} and the following, we define
 the baryon density ($B=n,p,\Lambda$)
\begin{equation}
  \rho_B(\br) = \sum_{i,\sigma}n_i|\phi_{i,B}(\br,\sigma)|^2,
\end{equation}
the kinetic energy density
\begin{equation}
  \tau_B(\br)= \sum_{i,\sigma}  n_i |\bnab\phi_{i,B}(\br,\sigma)|^2,
\end{equation}
and the spin density
\begin{equation}
 \bJ_B(\br) = -i \sum_{i,\sigma,\sigma'}n_i\phi^*_{i,B}(\br,\sigma) 
 \left[ \bnab \times\bsig  \phi_{i,B}(\br,\sigma')  \right],   
\end{equation}
where $\phi_{i,B}(\br,\sigma)$ is the wave function of the single-particle state, and $n_i$ is the corresponding occupation number, 
which is defined by  $n_i=v^2_i(2j+1)$. The  occupation probability $v^2_i$ of the single-particle state $i$ will be 
determined by either  BCS or the filling approximation depending on the model.  In Eq.~\eqref{eq:total-H-N}, the nucleon total densities are defined as $\rho_N=\rho_n+\rho_p$,  $\tau_N=\tau_n+\tau_p$,  and $\bJ_N=\bJ_n+\bJ_p$.

The hamiltonian density related with $\Lambda$ can be written as~\cite{Guleria2012NPA}
\begin{eqnarray}
\mathcal{H}_{\Lambda} &=& \frac{\hbar^2}{2m_{\Lambda}}\tau_{\Lambda}+
 t^{\Lambda}_0\left(1+ \half x^{\Lambda}_0\right)\rho_{\Lambda}\rho_N  
+   \frac{1}{4} \left( t^{\Lambda}_1 + t^{\Lambda}_2 \right)
    \left( \tau_{\Lambda}\rho_N +  \tau_{N}\rho_{\Lambda} \right) \nonumber \\
&& 
+\frac{1}{8}\left( 3t^{\Lambda}_1-t^{\Lambda}_2\right)   \bnab \rho_{\Lambda} \cdot \bnab \rho_N
+\frac{1}{2}W^{\Lambda}_0\left( \bnab\rho_{N} \cdot \bJ_{\Lambda}+ \bnab \rho_{\Lambda} \cdot \bJ_{N}\right)   \nonumber \\
&& 
+    \frac{3}{8} t^{\Lambda}_3 \left(  1+\half x_3\right) \rho^{\gamma+1}_N \rho_{\Lambda}.
\end{eqnarray}

As a first step, we assume the spherical symmetry for the hypernucleus, and the pairing correlation is not considered explicitly, but the filling approximation is adopted for the occupation probability $v_i^2$ from the bottom of potential to the Fermi energy in order.  The single-particle wave function for nucleons and $\Lambda$ can be written as
\begin{equation}
\phi_{i,B}(\br\sigma) = \frac{R_{i,B}(r)}{r}Y_{ljm}(\hat{\br}\sigma), ~~i=(nljm)~~ {\rm and} ~~ B=(n,p,\Lambda), 
\end{equation}
where $R_{i,B}(r)$ is the radial wave function, and $Y_{ljm}(\hat{\br}\sigma)$ is the vector spherical harmonics. 

To show the model-dependence of the calculation, we choose three Skyrme $NN$ interactions SIII~\cite{Beiner1975NPA}, SLy4~\cite{Chabanat1998NPA} and SkM*~\cite{Bartel1982NPA}, together with different Skyrme-type $\Lambda N$ interactions such as No. 1 in Ref. ~\cite{Yamamoto1988PTP} (labeled as 'YBZ1') fitted according to the hypernucleus data, No. 1 and 5 in Ref.~\cite{Lanskoy1997PRC} (labeled as 'LY1' and 'LY5') obtained by the $G-$matrix calculation from the one-boson-exchange potential.  
In particular, LY5 includes the $\Lambda N$ spin-orbit interaction with the strength $W^{\Lambda}_0=62$~MeV~fm$^5$.  However, we found the obtained spin-orbit splitting of the $1p$ states in $^{13}_{\Lambda}$C is too large compared to the experiment data $0.152$~MeV~\cite{Ajimura2001PRL}.  Therefore, we use a reduced value $W^{\Lambda}_0=4.7$~MeV~fm$^5$ instead (labeled as 'LY5r'), and obtain a realistic  spin-orbit splitting $0.155$~MeV of $1p$ states in $^{13}_{\Lambda}$C calculated with SkM*.  

The center of mass correction is considered simply by multiplying the factor ${1-m_N/(Am_N+m_{\Lambda})}$ and ${1-m_{\Lambda}/(Am_N+m_{\Lambda})}$ in front of the mass terms ${\hbar^2/2m_N}$ and ${\hbar^2/2m_{\Lambda}}$ respectively. 
The binding 
energy of $\Lambda$ particle can be calculated by 
\begin{equation}
  B_{\Lambda}=B^{\Lambda}_{A+1}-B_A ,
\end{equation}
where $B_A$ is the total binding energy of the nucleus with $A$ nucleons, and $B^{\Lambda}_{A+1}$ is the total binding energy of the hypernucleus with one additional $\Lambda$.  

\clearpage
\section{Results and Discussions}
\subsection{Hypernuclei of C isotopes}

We first discuss C isotopes since the spin-orbit splitting of hyperon states was observed only in $^{13}_{\Lambda}$C. 

\begin{figure}[!h]
	\centering
	\includegraphics[width=0.5\textwidth]{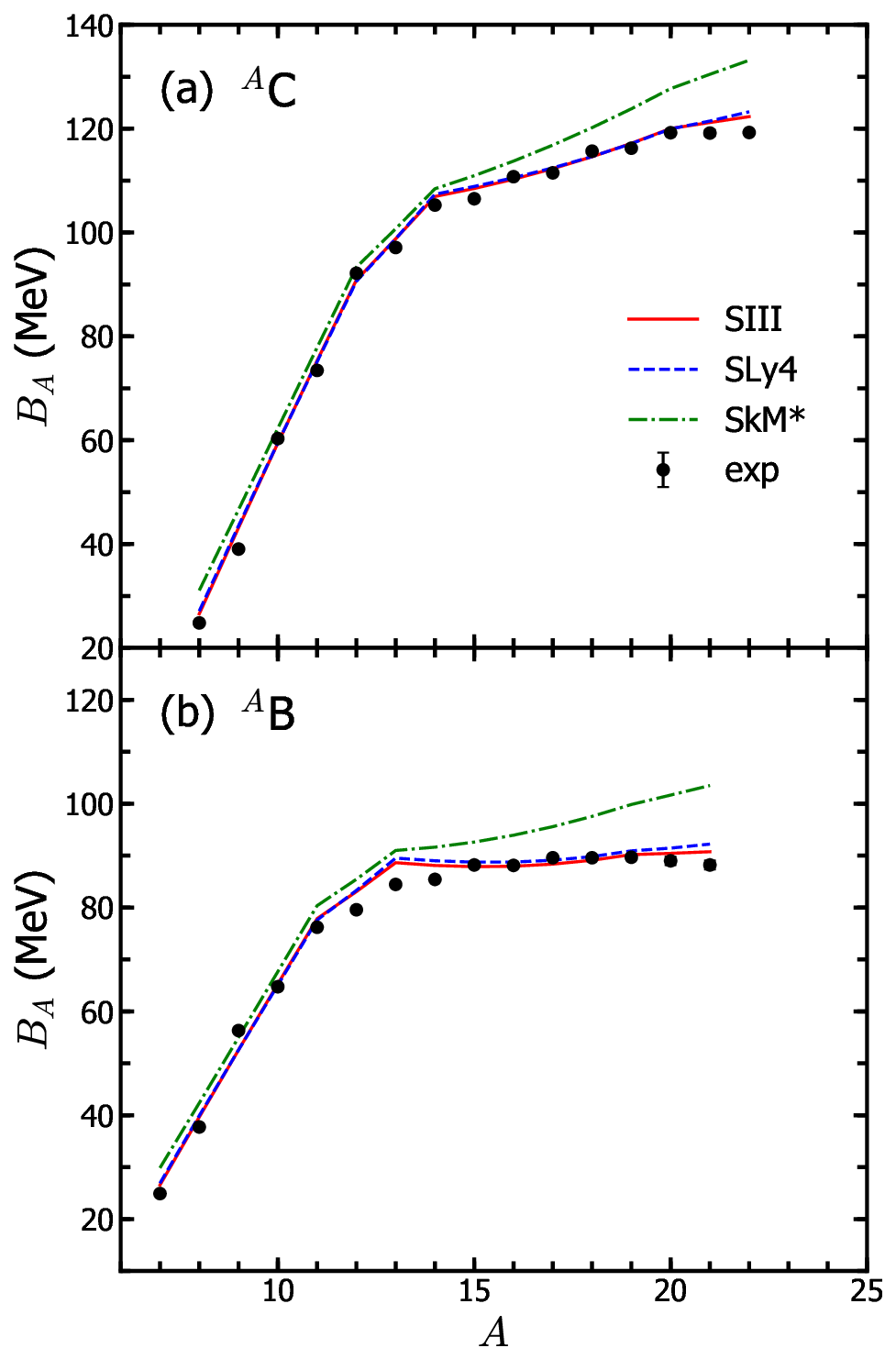}\\
	\caption{Total binding energies of (a) Carbon and (b) Boron isotopes with mass number $A$ calculated with different Skyrme $NN$ interactions: SIII, SLy4 and SkM*.  The experiment data~\cite{Wang2017AME} are also shown.}
	\label{fig:C-B-BE-S3-SLy4-SkMs}
\end{figure}

Without $\Lambda$ hyperon, the total binding energies of $^{8-22}$C calculated with Skyrme $NN$ interactions SIII, SLy4 and SkM* are shown in Fig.~\ref{fig:C-B-BE-S3-SLy4-SkMs}~(a).  The experiment data taken from Ref.~\cite{Wang2017AME} are also shown. One can see that, the results of SIII and SLy4 are quite consistent with the data, while SkM* provides more binding for the C isotopes with $A\geqq 15$.   The deformation effect might play a minor role here.  In the present model, the results of all these three $NN$ interactions show that the neutron drip line is $^{22}$C.  

\begin{figure}[!h]
	\centering
	\includegraphics[width=0.5\textwidth]{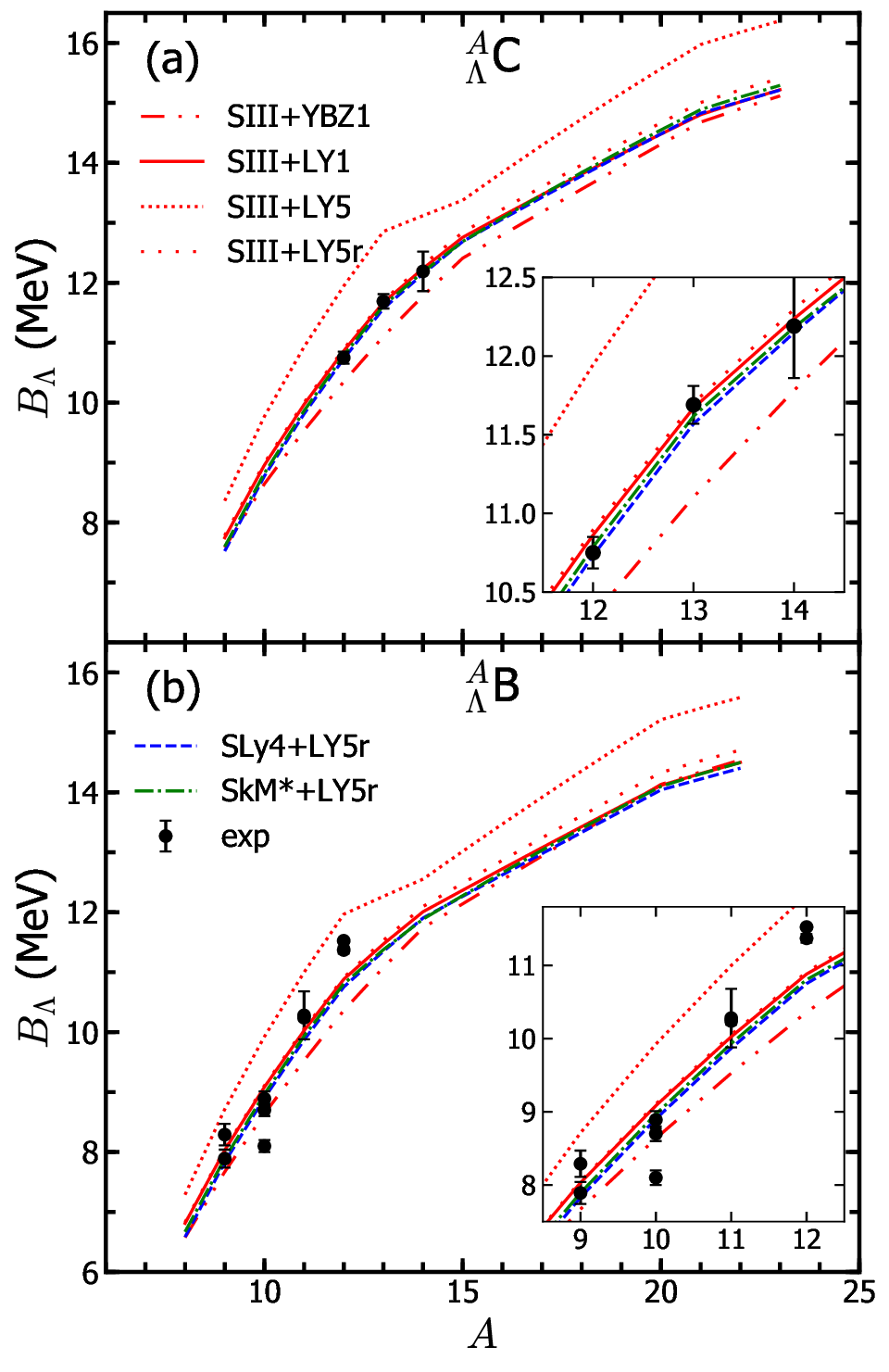}\\
	\caption{Lambda binding energy $B_{\Lambda}$ of the ground state of hypernucleus (a) $^A_{\Lambda}$C and (b) $^A_{\Lambda}$B calculated with Skyrme functionals for different $NN$ interactions: SIII, SLy4 and SkM*, and different $\Lambda N$ interactions: YBZ1, LY1, LY5, LY5r.  The results of SIII+LY1 (solid line) are almost identical to those of SIII+LY5r (loosely dotted line), Sly4+LY5r (dashed line), and SkM*+LY5r (dash-dotted line), which are enlarged in the insets for $^{12-14}_{\Lambda}$C and $^{9-12}_{\Lambda}$B respectively.
	The experiment data~\cite{Pile1991PRL,CANTWELL-1974,Juric1973NPB,Hasegawa,Davis2005NPA,Tang-2014,Botta2017NPA} are also shown.}
	\label{fig:CLam-BLam-S3-SLy4-SkMs}
\end{figure}

Adding one $\Lambda$ hyperon inside the C isotopes, the $\Lambda$ binding energies of the ground state $1s$ calculated with Skyrme-type $\Lambda N$ interactions YBZ1, LY1, LY5 and LY5r are shown in Fig.~\ref{fig:CLam-BLam-S3-SLy4-SkMs}~(a).  
The experiment data are taken from Ref.~\cite{Pile1991PRL} for $^{12}_{\Lambda}$C, and Ref.~\cite{CANTWELL-1974} for $^{13,14}_{\Lambda}$C.  With the  $NN$ interaction SIII, 
$\Lambda N$ interaction LY1 gives the nice prediction for the $\Lambda$ binding energy, while YBZ1 leads to a bit less binding and the original LY5 obvious over-binding comparing to the available data.  With the reduced spin-orbit strength $W^{\Lambda}_0$, LY5r could give quite consistent predictions for the $\Lambda$ binding energy using different $NN$ interactions, which also agree very well with the available data.

The $\Lambda$ binding energies of the $1p$ states calculated with the same $NN$ and $\Lambda N$ interactions are shown in Table~\ref{tab:1p-BE-Sk3-SLy4-SkMs}.  In YBZ1 and LY1, there is no $\Lambda N$ spin-orbit interaction.  While in LY5 and LY5r, with the $\Lambda N$ spin-orbit interaction, the first and second lines list the binding energies of $1p_{1/2}$ and $1p_{3/2}$ states respectively.  One could see that, in $^{12}_{\Lambda}$C, most of the $1p$ states are unbound with respect to the $^{11}$C$+\Lambda$ threshold, since their binding energies $B_{\Lambda}$ are negative.  In $^{13}_{\Lambda}$C, most of the results show the weakly bound $1p$ states.  With the original spin-orbit strength $W^{\Lambda}_0=62$~MeV fm$^5$, the interactions SIII+LY5 leads to the spin-orbit splitting nearly $2$~MeV between $1p_{1/2}$ and $1p_{3/2}$ states.  However, the experiment data~\cite{Kohri-2002} showed this splitting is only $0.152$~MeV.  
	With the reduced value $W^{\Lambda}_0=4.7$~MeV fm$^5$ in   LY5r, different $NN$ interactions SIII, SLy4 and SkM* obtain the consistent $1p$ spin-orbit splittings $0.153$~MeV, $0.149$~MeV, and $0.155$ MeV respectively in $^{13}_{\Lambda}$C. 
Besides, in Ref. \cite{Kohri-2002},  the excitation energies of 
$\Lambda( 1p_{1/2})$ and $\Lambda( 1p_{3/2 })$ states were observed as $E_x=10.982\pm0.031$(stat)$\pm$0.056(syst) MeV and $E_x=10.830\pm0.031$(stat)$\pm$0.056(syst) MeV,
respectively.  The values calculated with SkM*+LY5r are $E_x$=11.344 and 11.190 MeV for $\Lambda(1p_{1/2})$ and $\Lambda (1p_{3/2})$ states, which show reasonable agreement with the experiment data.  		
	 With more neutrons, the $1p$ states becomes more deeply bound.  But the spin-orbit splittings are almost constant.  Moreover, with the same $\Lambda N$ interaction LY5r, the $\Lambda$ binding energies and spin-orbit splittings of $1p$ states calculated with different $NN$ interactions are  consistent with each other in heavier C hypernuclei.

The above investigations show that the $\Lambda$ binding energy is mainly determined by the $\Lambda N$ interaction, almost independent on the $NN$ interaction.  
In the following, we will take the results calculated with the $NN$ interaction SkM* and $\Lambda N$ interaction LY5r as examples to discuss the possible $\Lambda$ halo states in C isotopes.  

\begin{table}[htbp]
	\renewcommand\arraystretch{0.75}
	\centering
	\caption{Lambda binding energies of the $1p$ states in $^A_{\Lambda}$C calculated with different $NN$ (SIII, SLy4, SkM*) and $\Lambda N$ (YBZ1, LY1, LY5, LY5r) effective interactions.  The $\Lambda N$ spin-orbit interaction is included in LY5 and LY5r, where the first and second lines show the binding energies of $1p_{1/2}$ and $1p_{3/2}$ states respectively.   }
	\begin{tabular}{rrrrrrr}
		\hline\hline
		Nucleus & \multicolumn{1}{c}{SIII+YBZ1} & \multicolumn{1}{c}{SIII+LY1} & \multicolumn{1}{c}{SIII+LY5} & \multicolumn{1}{c}{SIII+LY5r} & \multicolumn{1}{c}{SLy4+LY5r} & \multicolumn{1}{c}{SkM*+LY5r}  \\ \hline
		$^{12}_{\Lambda}$C    & -0.961  & -0.521  & -1.367  & -0.385  & -0.329  & -0.379  \\
		&               &              &  0.461  & -0.243  & -0.194  &  -0.239  \\ \hline
		$^{13}_{\Lambda}$C    & -0.305  & 0.187   & -0.758  & 0.312    & 0.324   & 0.273  \\
		&               &              & 1.226    & 0.465   &  0.473  &  0.428   \\ \hline
		$^{14}_{\Lambda}$C    & 0.439   & 0.917   & -0.041   & 1.049   & 1.044   & 1.010  \\
		&               &              & 1.912    & 1.199   &  1.190  & 1.160  \\ \hline
		$^{15}_{\Lambda}$C    & 1.155   & 1.606   & 0.647    & 1.741   & 1.723    & 1.697  \\
		&               &              & 2.554    & 1.888   & 1.866   &  1.842 \\ \hline
		$^{16}_{\Lambda}$C    & 1.649   & 2.095   & 1.193    &  2.241   & 2.207   & 2.187  \\
		&               &              &  3.088   & 2.386    & 2.348   &  2.331  \\ \hline
		$^{17}_{\Lambda}$C    & 2.140    & 2.575   & 1.731   & 2.729    & 2.680   &  2.666  \\
		&               &              & 3.608   & 2.872    & 2.819   &   2.809  \\ \hline
		$^{18}_{\Lambda}$C    & 2.627    & 3.043    & 2.259  & 3.204    & 3.142   & 3.133  \\
		&               &              & 4.115   &3.345     &  3.280  & 3.274    \\ \hline
		$^{19}_{\Lambda}$C    & 3.108     & 3.501  & 2.775    & 3.667    & 3.594  & 3.587  \\
		&               &              & 4.607    & 3.806    &  3.730  & 3.727 \\ \hline
		$^{20}_{\Lambda}$C     & 3.583    & 3.947   & 3.278    & 4.118    & 4.037   & 4.028  \\
		&                &              & 5.086   &  4.254   &  4.172  &  4.167 \\ \hline
		$^{21}_{\Lambda}$C    & 4.051     & 4.383    & 3.770   & 4.556    & 4.472    & 4.457  \\
		&                &              &  5.553   & 4.691    & 4.605   &  4.595 \\ \hline
		$^{22}_{\Lambda}$C     & 4.331    & 4.643    & 4.021    & 4.833    & 4.736    & 4.742  \\
		&                &              & 5.820     &  4.969   & 4.869     &  4.881  \\ \hline
		$^{23}_{\Lambda}$C     & 4.586     & 4.880   & 4.250     & 5.086    & 4.991   & 5.000  \\  
		&               &              & 6.066     & 5.223     & 5.127  & 5.140   \\ \hline
	\end{tabular}%
	  \label{tab:1p-BE-Sk3-SLy4-SkMs}%
\end{table}%

  \begin{table}[!hbt] 
	\renewcommand\arraystretch{0.6}
	\centering 
	\caption{Properties of single-$\Lambda$ states in hypernucleus $^{A}_{\Lambda}$C calculated with the Skyrme $NN$ interaction SkM* and $\Lambda N$ interaction LY5r: single-particle energy $e_{\rm s.p.}$, the rms radius $r^{\Lambda}_{\rm rms}$ of the corresponding singe-particle state, $B(E1)$ value of the transition from the excited $\Lambda(1p)$-state to the ground $\Lambda(1s)$-state. 
	 \label{C-hyperon}}
	\begin{tabular}{ccrrr} 
		\hline \hline 
		Nucleus  & $\Lambda (nlj)$ & $e_{\rm s.p.}$ (MeV) & $r^{\Lambda}_{\rm rms}$ (fm) & $B(E1)$  ($e^2$fm$^2$)  \\  \hline
		$^{9}_{\Lambda}$C   &   $1s_{1/2}$  & $-9.478$      & $2.160$  &      \\ \hline
		$^{10}_{\Lambda}$C &   $1s_{1/2}$  & $-10.662$      &  $2.141$    &         \\ \hline
		$^{11}_{\Lambda}$C &   $1s_{1/2}$  & $-11.615$      &  $2.136$    &         \\ \hline
		$^{12}_{\Lambda}$C &	$1s_{1/2}$  &	$-12.433$      &	 $2.139$   &  	 \\ 
		&    $1p_{1/2}$  &   $-1.228$          &  $3.679$   &	$1.176\times 10^{-1}$  \\
		&    $1p_{3/2}$  &   $-1.367$          &  $3.604$   &	$1.186\times 10^{-1}$  \\ 
		\hline
		$^{13}_{\Lambda}$C &	$1s_{1/2}$	&   $-13.156$     &  $2.144$    &  \\
		&    $1p_{1/2}$  &   $-1.782$          &  $3.464$    &	  $1.030\times 10^{-1}$\\
		&    $1p_{3/2}$  &   $ -1.936$         &  $3.410$    &	$1.036\times 10^{-1}$   \\ \hline                                
		$^{14}_{\Lambda}$C &	$1s_{1/2}$  &	$-13.563$    & 	$2.172$ &  \\
		&     $1p_{1/2}$  &	$-2.357$        & 	$3.355$  & 	$9.264\times 10^{-2}$  \\ 
		&     $1p_{3/2}$  &	$-2.506$        &	$3.317$  & 	$9.297\times 10^{-2}$ \\ \hline	        
		$^{15}_{\Lambda}$C &	$1s_{1/2}$	&  $-13.941$    & 	$2.199$  &	 \\
		&    $1p_{1/2}$  &	$-2.911$     & 	$3.287$   & 	$8.367\times 10^{-2}$   \\
		&    $1p_{3/2}$  &	$-3.055$    & 	$3.259$   & 	$8.385\times 10^{-2}$	\\  
		\hline    
		$^{16}_{\Lambda}$C &	$1s_{1/2}$  &  $-14.292$    &     $2.218$  & \\	
		&    $1p_{1/2}$  &   $-3.357$ 	    &     $3.252$  &  $7.524\times 10^{-2}$  \\
		&    $1p_{3/2}$  &   $-3.500$ 	    &     $3.228$  &  $7.537\times 10^{-2}$  \\	\hline                                  
		$^{17}_{\Lambda}$C &	$1s_{1/2}$	&   $-14.633$   &    $2.236$ 	 &  \\
		&    $1p_{1/2}$  &  $-3.792$ 	  &    $3.226$   &	$6.806\times 10^{-2}$  \\
		&    $1p_{3/2}$  &  $-3.935$      &    $3.206$   &	$6.814\times 10^{-2}$  \\ 
		\hline
		$^{18}_{\Lambda}$C &	$1s_{1/2}$	&  $-14.962$    &    $2.254$ 	 & \\
		&    $1p_{1/2}$  &   $-4.216$ 	&    $3.207$   &	$6.188\times 10^{-2}$  \\
		&    $1p_{3/2}$  &   $-4.357$ 	&    $3.189$   & $6.194\times 10^{-2}$ \\
		\hline      
		$^{19}_{\Lambda}$C &	$1s_{1/2}$	&  $-15.281$    &    $2.270$ 	 &  \\
		&    $1p_{1/2}$  &  $-4.629$ 	    &	$3.192$   &	$5.653\times 10^{-2}$  \\
		&    $1p_{3/2}$  &  $-4.769$ 	    &    $3.177$   & $5.657\times 10^{-2}$  \\
		\hline
		$^{20}_{\Lambda}$C &	$1s_{1/2}$	 &  $-15.590$   &    $2.286$ 	 & \\
		&    $1p_{1/2}$	&  $-5.031$ 	&    $3.182$   & $5.188\times 10^{-2}$ \\
		&    $1p_{3/2}$	&  $-5.170$ 	&    $3.168$   & $5.190\times 10^{-2}$ \\ 
		\hline
		$^{21}_{\Lambda}$C &	$1s_{1/2}$	 &  $-15.890$   &    $2.302$ 	 &  \\
		&    $1p_{1/2}$  &   $-5.422$     &    $3.174$   & $4.780\times 10^{-2}$ \\
		&    $1p_{3/2}$  &   $-5.559$ 	  &    $3.162$   & $4.782\times 10^{-2}$     \\ \hline
		$^{22}_{\Lambda}$C &	$1s_{1/2}$	 &  $-16.038$   &    $2.315$ 	 & \\
		&    $1p_{1/2}$  &   $-5.648$ 	&	$3.191$   & $4.405\times 10^{-2}$ \\
		&    $1p_{3/2}$  &   $-5.787$   &    $3.178$   &  $4.407\times 10^{-2}$	\\ \hline
		$^{23}_{\Lambda}$C &	$1s_{1/2}$	 &   $-16.176$  &    $2.326$   &  \\ 	
		&    $1p_{1/2}$	&   $-5.853$ 	&    $3.208$   &	$4.070\times 10^{-2}$  \\
		&    $1p_{3/2}$	&   $-5.992$    &    $3.195$   &	$4.072\times 10^{-2}$	\\ 
		\hline \hline
	\end{tabular}
\end{table}

\begin{table}[htbp]
	\centering	
	\renewcommand\arraystretch{0.6}
	\caption{ 
		The calculated mass rms radius $r^{\rm A}_{\rm rms}$ of isotopes $^A$C, the corresponding experiment data $r^{\rm A}_{\rm rms}({\rm exp})$ taken from Refs.~\cite{Ozawa2001,Togano2016}, and
		the calculated mass rms radius of the core $r^{\rm core A}_{\rm  rms}$ in hypernucleus $^{A+1}_{\Lambda}$C.  }
	\begin{tabular}{rcccc}
		\hline\hline
		\multicolumn{1}{l}{Nucleus} & \multicolumn{1}{l}{$r^{\rm A}_{\rm rms}$ (fm) } &$r^{\rm A}_{\rm rms}({\rm exp})$ (fm) &$\Lambda(nlj)$ & \multicolumn{1}{l}{$r^{\rm core A}_{\rm  rms}$ (fm)} \\  \hline
		\multicolumn{1}{l}{$^{8}$C}    & $2.5573$ &  & $1s_{1/2}$  &   $2.5020$  \\  \hline  
		\multicolumn{1}{l}{$^{9}$C}   &  $2.4408$ &&  $1s_{1/2}$  & $2.4120$  \\ \hline
		\multicolumn{1}{l}{$^{10}$C} &  $2.4098$  & & $1s_{1/2}$ & $2.3902$  \\ \hline 
		\multicolumn{1}{l}{$^{11}$C} & $2.4094$  & &$1s_{1/2}$ & $2.3942$  \\
		      &       && $1p_{1/2}$ & $2.4169$  \\
	          &       && $1p_{3/2}$ & $2.4158$  \\ \hline
		\multicolumn{1}{l}{$^{12}$C}  & $2.4228$ &2.35$\pm$0.02 & $1s_{1/2}$ & $2.4103$  \\
		      &      & & $1p_{1/2}$ & $2.4290$  \\
		      &       && $1p_{3/2}$ & $2.4280$  \\  \hline
		\multicolumn{1}{l}{$^{13}$C}   & $2.5095$ &2.28$\pm$ 0.04& $1s_{1/2}$ & $2.4943$  \\
	          &      & & $1p_{1/2}$ & $2.5125$  \\
	          &       && $1p_{3/2}$ & $2.5116$  \\  \hline
		\multicolumn{1}{l}{$^{14}$C}  & $2.5860$  & 2.30$\pm$0.07 &$1s_{1/2}$ & $2.5690$  \\
		       &       && $1p_{1/2}$ & $2.5865$  \\
		       &       & & $1p_{3/2}$ & $2.5856$  \\  \hline
		\multicolumn{1}{l}{$^{15}$C} & $2.6570$  & 2.50$\pm$0.08 &$1s_{1/2}$ & $2.6385$  \\
		       &       & &$1p_{1/2}$ & $2.6554$  \\
		       &       & &$1p_{3/2}$ & $2.6545$  \\  \hline
		\multicolumn{1}{l}{$^{16}$C}  & $2.7193$  & 2.70$\pm$0.03& $1s_{1/2}$ & $2.6999$  \\
		       &       && $1p_{1/2}$ & $2.7160$  \\
		       &       & &$1p_{3/2}$ & $2.7152$  \\  \hline
		\multicolumn{1}{l}{$^{17}$C} & $2.7747$  &2.72$\pm$0.03   & $1s_{1/2}$ & $2.7545$  \\
		       &       && $1p_{1/2}$ & $2.7700$  \\
		       &       & &$1p_{3/2}$ & $2.7692$  \\  \hline
		\multicolumn{1}{l}{$^{18}$C}   & $2.8243$  & 2.82$\pm$0.04    &$1s_{1/2}$ & $2.8037$  \\
		       &       && $1p_{1/2}$ & $2.8186$  \\
		       &       & &$1p_{3/2}$ & $2.8179$  \\  \hline
		\multicolumn{1}{l}{$^{19}$C}    & $2.8692$ &  3.13$\pm$0.07& $1s_{1/2}$ & $2.8484$  \\
		       &       && $1p_{1/2}$ & $2.8628$  \\
		       &       & &$1p_{3/2}$ & $2.8620$  \\  \hline
		\multicolumn{1}{l}{$^{20}$C}   & $2.9102$ & 2.98$\pm$0.05 & $1s_{1/2}$ & $2.8894$  \\
		       &       && $1p_{1/2}$ & $2.9032$  \\
		       &       & &$1p_{3/2}$ & $2.9025$  \\  \hline
		\multicolumn{1}{l}{$^{21}$C}   & $3.0054$  &  &$1s_{1/2}$ & $2.9833$  \\
		       &       & &$1p_{1/2}$ & $2.9952$  \\
		       &       & &$1p_{3/2}$ & $2.9944$  \\  \hline
		\multicolumn{1}{l}{$^{22}$C}   & $3.0995$  & 3.44$\pm$0.08 &$1s_{1/2}$ & $3.0762$  \\
		       &       & &$1p_{1/2}$ & $3.0865$  \\
		       &       & &$1p_{3/2}$ & $3.0858$  \\  \hline
	\end{tabular}%
	\label{C-core}%
\end{table}%

\begin{figure}[!h]
	\centering
	\includegraphics[width=0.5\textwidth]{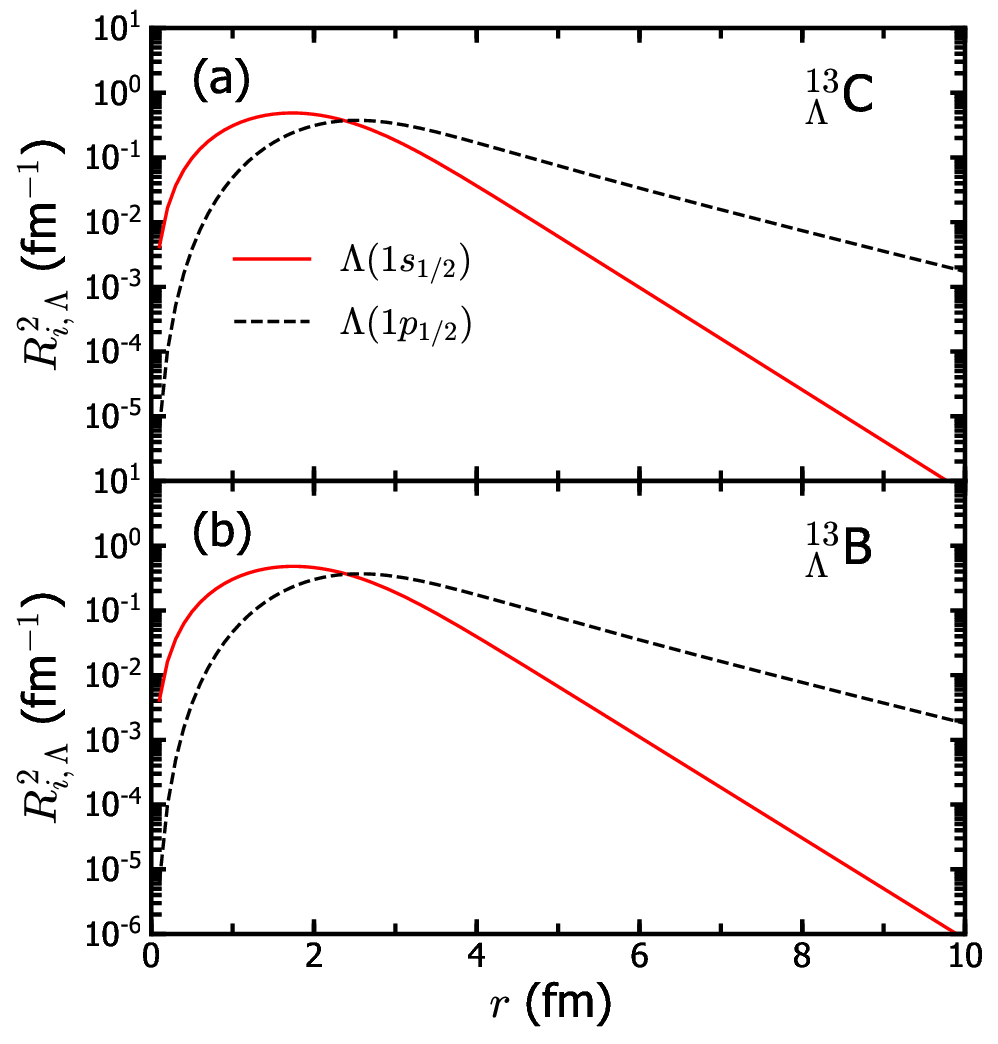}\\
	\caption{The square of single-$\Lambda$ wave function $R^2_{i,\Lambda}$ of $\Lambda(1s_{1/2})$ and $\Lambda(1p_{1/2})$ states in the hypernucleus (a) $^{13}_{\Lambda}$C and (b) $^{13}_{\Lambda}$B respectively. \label{C13-B13-spwf}}
\end{figure}

The HF single-particle energies and rms radii of $\Lambda(1s)$- and $\Lambda(1p)$-orbits in C isotopes are listed in Table \ref{C-hyperon}.  The $\Lambda(1p)$ states in $^{12-14}_\Lambda$C are quasi-bound (resonant) or loosely-bound states as shown in Table~\ref{tab:1p-BE-Sk3-SLy4-SkMs}.  Especially, their rms radii show a peculiar halo nature similar to the  halo state in nuclei such as $^{11}$Li and $^{11}$Be.  The wave functions of $\Lambda(1s_{1/2})$- and $\Lambda(1p_{1/2})$-orbits in $^{13}_\Lambda$C are plotted in Fig.~\ref{C13-B13-spwf}~(a).  The enhancement of rms radii of $\Lambda(1p)$-orbit is about 60\% compared with the $\Lambda(1s)$-orbit.   Thus we can conclude to find the $\Lambda(1p)$ halo state in  $^{13}_\Lambda$C.  For $^{12}_{\Lambda}$C and $^{14}_{\Lambda}$C hypernuclei,  the $\Lambda(1p)$ states also have small binding energies and show the similar halo structure to that of $^{13}_\Lambda$C.  

The matter rms radii $r^{\rm  A}_{\rm  rms}$ of C isotopes are tabulated in Table~\ref{C-core}.  
The listed mass radii of C isotopes are observed by heavy-ion reactions~\cite{Ozawa2001,Togano2016}.  The  calculated results reproduce reasonably well the experiment values
 except the neutron halo nuclei $^{19}$C and $^{22}$C.   
  The rms radii of the cores of corresponding hypernuclei are  also listed as $r^{\rm core A}_{\rm  rms}$.   In comparison between 
  $r^{\rm  A}_{\rm  rms}$ and $r^{\rm core A}_{\rm  rms}$, we can find shrinkage or expansion effect of the core nucleus in hypernucleus.  
  For $\Lambda(1s)$ hyperon case, we can see small shrinkage effect of the core,  $0.05-0.02$ fm,  from light to heavy C isotopes.  
  For $\Lambda(1p)$ hyperon case,  it is interesting to see an expansion effect of the core for nuclei $A\leq 13$, 
  but quantitatively it is even smaller than the shrinkage effect of $\Lambda(1s)$
   hyperon in the same nucleus.

\subsection{Hypernuclei of B isotopes}
  \begin{table}[!hbt]
	\renewcommand\arraystretch{0.6}
	\centering 
	\caption{
		Properties of single-$\Lambda$ states in hypernucleus $^{A}_{\Lambda}$B calculated with the Skyrme $NN$ interaction SkM* and $\Lambda N$ interaction LY5r: single-particle energy $e_{\rm s.p.}$, binding energy $B_{\Lambda}$, rms radius $r^{\Lambda}_{\rm rms}$ of the corresponding single-particle state, 
		$B(E1)$ value of the transition from the excited $\Lambda(1p)$-state to the ground $\Lambda(1s)$-state. 
		 \label{B-hyperon}}
	\begin{tabular}{ccrrrr} 
		\hline \hline 
		Nucleus  & $\Lambda(nlj)$ & $e_{\rm s.p.}$ (MeV) & $B_{\Lambda}$ (MeV) & $r^{\Lambda}_{\rm rms}$ (fm) & $B(E1)$  ($e^2$fm$^2$) \\  \hline
		$^{8}_{\Lambda}$B    &	$1s_{1/2}$    & 	 $-8.750$  &  $6.670$     &  $2.148$   &  	\\  \hline
		$^{9}_{\Lambda}$B    &	$1s_{1/2}$    & 	 $-9.917$  &  $7.892$     &  $2.132$   &  \\  \hline
		$^{10}_{\Lambda}$B  &	$1s_{1/2}$   & 	 $-10.877$  &  $8.968$     &  $2.128$   &  \\ \hline
		$^{11}_{\Lambda}$B  &	$1s_{1/2}$   & 	 $-11.712$  &  $9.932$     &  $2.131$   &   \\ \hline
		$^{12}_{\Lambda}$B &	$1s_{1/2}$   & 	 $-12.457$  &  $10.805$     &  $2.137$   &  \\
		&    $1p_{1/2}$  &   $-1.229$     &  $-0.386$     &   $3.674$   &	$8.1524\times 10^{-2}$ \\
		&    $1p_{3/2}$  &   $-1.370$     &  $-0.245$     &   $3.599$   & $8.2226\times 10^{-2}$ \\ 
		\hline
		$^{13}_{\Lambda}$B &	$1s_{1/2}$   & 	 $-12.843$  &  $11.375$     &  $2.168$    &  	\\ 
		&    $1p_{1/2}$  &   $-1.787$     &  $0.364$       &   $3.503$   & $7.3314\times 10^{-2}$  \\
		&    $1p_{3/2}$  &   $-1.925$     &  $0.502$       &   $3.454$   & $7.3684\times 10^{-2}$  \\ 	  
		\hline
		$^{14}_{\Lambda}$B &	$1s_{1/2}$   & 	 $-13.205$  &  $11.885$     &  $2.198$    &  	\\ 
		&    $1p_{1/2}$  &   $-2.331$     &  $1.061$       &   $3.402$   & $6.6020\times 10^{-2}$  \\
		&    $1p_{3/2}$  &   $-2.465$     &  $1.195$       &   $3.367$   & $6.6216\times 10^{-2}$  \\ 
		\hline  
		$^{15}_{\Lambda}$B &	$1s_{1/2}$   & 	 $-13.544$  &  $12.277$     &   $2.218$   &  	\\ 
		&    $1p_{1/2}$  &   $-2.765$     &  $1.549$       &   $3.352$   & $5.9055\times 10^{-2}$  \\
		&    $1p_{3/2}$  &   $-2.898$     &  $1.682$       &   $3.323$   & $5.9190\times 10^{-2}$  \\  \hline
		$^{16}_{\Lambda}$B  &	$1s_{1/2}$   & 	 $-13.876$  &  $12.660$     &  $2.237 $   &  	 \\ 
		&    $1p_{1/2}$  &   $-3.193$    &  $2.028$       &   $3.315$   & $5.3143\times 10^{-2}$  \\
		&    $1p_{3/2}$  &   $-3.325$    &  $2.160$       &   $3.291$   & $5.3233\times 10^{-2}$  \\ 	 \hline   
		$^{17}_{\Lambda}$B  &	$1s_{1/2}$   & 	 $-14.203$  &  $13.034$     &  $2.255$    & 	\\ 
		&    $1p_{1/2}$  &   $-3.613$     &  $2.497$       &   $3.287$   & $4.8090\times 10^{-2}$  \\
		&    $1p_{3/2}$  &   $-3.744$     &  $2.628$       &   $3.266$   & $4.8151\times 10^{-2}$  \\ 	\hline  
		$^{18}_{\Lambda}$B  &	$1s_{1/2}$   & 	 $-14.522$  &  $13.399$     &  $2.273$    &  	 \\ 
		&    $1p_{1/2}$  &   $-4.026$     &  $2.956$       &   $3.265$   & $4.3743\times 10^{-2}$  \\
		&    $1p_{3/2}$  &   $-4.156$     &  $3.086$       &   $3.247$   & $4.3784\times 10^{-2}$  \\ 	\hline        
		$^{19}_{\Lambda}$B &	$1s_{1/2}$   & 	 $-14.834$  &  $13.754$     &  $2.290$    & 	\\ 
		&    $1p_{1/2}$  &   $-4.430$     &  $3.403$       &   $3.249$   & $3.9977\times 10^{-2}$  \\
		&    $1p_{3/2}$  &   $-4.559$     &  $3.532$       &   $3.233$   & $4.0005\times 10^{-2}$  \\ 	\hline        
		$^{20}_{\Lambda}$B &	$1s_{1/2}$   & 	 $-15.138$  &  $14.099$     &  $2.306$    &   \\ 
		&    $1p_{1/2}$  &   $-4.825$     &  $3.839$       &   $3.236$   & $3.6695\times 10^{-2}$ \\
		&    $1p_{3/2}$  &   $-4.952$     &  $3.966$       &   $3.222$   & $3.6713\times 10^{-2}$  \\ 	\hline        
		$^{21}_{\Lambda}$B &	$1s_{1/2}$   & 	 $-15.276$  &  $14.306$     &  $2.319$    &   \\ 
		&    $1p_{1/2}$  &   $-5.034$     &  $4.112$       &   $3.254$   & $3.3640\times 10^{-2}$  \\
		&    $1p_{3/2}$  &   $-5.163$     &  $4.241$       &   $3.239$   & $3.3659\times 10^{-2}$  \\ 	\hline          
		$^{22}_{\Lambda}$B &	$1s_{1/2}$   & 	 $-15.406$  &  $14.497$     &  $2.330$    &  	 \\ 
		&    $1p_{1/2}$  &   $-5.224$     &  $4.360$       &   $3.271$   & $3.0932\times 10^{-2}$  \\
		&    $1p_{3/2}$  &   $-5.353$     &  $4.489$       &   $3.257$   & $3.0952\times 10^{-2}$  \\ 	\hline \hline                                                                                                                                                                          		                                                       
	\end{tabular}
\end{table}

The same calculations are also done for the hypernuclei of B isotopes.  First, the total binding energies of $^{7-21}$B without hyperons calculated with different Skyrme $NN$ interactions: SIII, SLy4 and SkM* are shown in Fig.~\ref{fig:C-B-BE-S3-SLy4-SkMs} (b), comparing with the experiment data~\cite{Wang2017AME}.  Similar with the results of C isotopes, SkM* provides more binding than SLy4 and SIII for $A>13$.  Although the spin-spin interaction is missing in the present Skyrme energy density functional which might play an important role in odd-even or odd-odd nuclei, most of the present results are consistent with the experiment data except $^{12-14}$B.  Adding one $\Lambda$ hyperon inside, the $\Lambda$ binding energies of the ground state $1s$ in the B hypernuclei calculated with different $\Lambda N$ and $NN$ interactions are shown in Fig.~\ref{fig:CLam-BLam-S3-SLy4-SkMs}~(b).  The experiment data are taken from Refs.~\cite{Juric1973NPB,Hasegawa,Davis2005NPA,Tang-2014,Botta2017NPA}.  Similar with the C hypernuclei, all the interaction combinations give the consistent $\Lambda$ binding energies except SIII+LY5, which makes the $\Lambda$ hyperon over-bind.  It is interesting to find that, although the $\Lambda N$ interaction `LY5r' is adjusted to the experiment data of $^{13}_{\Lambda}$C, the calculated results for B hypernuclei are also in reasonable agreement with the available experiment data, while there are some uncertainties in the experiment data.  The reasonable agreement between the calculated and experimental results of $B_\Lambda$ in Figs.~\ref{fig:C-B-BE-S3-SLy4-SkMs} and \ref{fig:CLam-BLam-S3-SLy4-SkMs} ensures the applicability of the present $\Lambda N$ interaction to  a wide mass region of hypernuclei, at least to most of  $p$-shell hypernuclei.

The $\Lambda$ single-particle energies, binding energies, and the rms radius of $1s$ and $1p$ states calculated with Skyrme $NN$ interaction SkM* and $\Lambda N$ interaction LY5r are listed in Table~\ref{B-hyperon}.  
 The potential depth is becoming deeper for heavier isotopes and the binding energy of $\Lambda(1s_{1/2})$-state increases from 8.97 MeV in $^{10}_{\Lambda}$B to 14.50 MeV in $^{22}_{\Lambda}$B.
The halo structure of $1p$-orbits can be also seen in light B isotopes, especially in $^{12}_{\Lambda}$B and $^{13}_{\Lambda}$B.
The wave functions of $\Lambda(1s_{1/2})$- and $\Lambda(1p_{1/2})$-orbits in $^{13}_\Lambda$B  are drawn in Fig. \ref{C13-B13-spwf}~(b).   The wave functions in $^{13}_\Lambda$B are essentially 
identical to those of  $^{13}_\Lambda$C.
The spin-orbit splittings in B isotopes show a similar feature to that in C isotopes; 
${\Delta \varepsilon (\Lambda(1p_{1/2})-\Lambda(1p_{3/2}))=0.138 {\rm MeV}}$ 
in $^{13}_\Lambda$B and ${\Delta \varepsilon (\Lambda( 1p_{1/2})-\Lambda( 1p_{3/2}))=0.129}$ MeV for a heavier isotope $^{22}_\Lambda$B.
Two $\Lambda(1p)$ states were also observed in \cite{Tang-2014},  as $J^\pi=(1_1^+ ~~{\rm or} ~~2_1^+)$ and $(2_2^+ ~~ {\rm or} ~~  3_1^+)$ states, which are considered as coupling states of
3/2$^-$ ground state of $^{11}$B and $\Lambda( 1p_{3/2})$ or  $\Lambda( 1p_{1/2})$ states.  Since the spin-spin interaction of $\Lambda N$ is not included in the present 
HF calculations, we can not predict precisely the energy splitting of $1^+, 2^+$ and $3^+$ states.  However, the HF excitation energies of $\Lambda(1p)$ states $E_x\sim11.1$ MeV are reasonable compared with the experiment data  $E_x$(exp)=$10.24\pm 0.05$ and $10.99\pm0.03$ MeV
for $J^\pi=(1_1^+ ~~{\rm or} ~~2_1^+)$ and $(2_2^+ ~~ {\rm or} ~~  3_1^+)$ states, respectively.


\subsection{Electic dipole transition in Hypernuclei}
We  will study the electric dipole transition between hyperon $1p$- and $1s$-state.  Electromagnetic transitions may provide precise information of 
hyperon wave functions in quantitative manner.  
Suppose the hypernucleus is initially in the excited state, e.g., $\Lambda$ is in the $1p$ orbit, 
it will decay to the ground state $1s$ orbit.  This $E1$ transition has the reduced transition probability~\cite{Ring2004Book}
\begin{equation}
B(E1; J_i \rightarrow J_f) = \frac{3e_{\Lambda}^2}{4\pi} \left< f | r | i \right>^2(2j_f+1)
\left( 
\begin{array}{ccc}
	j_f & 1  & j_i \\
	-\frac{1}{2} & 0 & \frac{1}{2}
\end{array} 
\right)^2,  
\label{eq:Lambda-sp-BE1}
\end{equation}
where $e_{\Lambda}$ is the effective charge for $\Lambda$ hyperon and  the 
integration $\left<  f | r | i \right> $ can be calculated by the radial wave functions of the initial and final single-$\Lambda$ state as
\begin{equation}
\left<  f | r | i \right> = \int ^\infty_0 R_{f,\Lambda}(r) r R_{i,\Lambda}(r)dr.  
\end{equation}
Since hyperons $\Lambda$ have no electric charges, the effective charge in Eq. (\ref{eq:Lambda-sp-BE1}) is given as 
\begin{equation}
e^{(E1)}_{\Lambda} = - Z M_{\Lambda} e /(AM_N+M_{\Lambda}), 
\label{eq:Lambda-BE1-effelam}
\end{equation}
due to the recoil of the core nucleus~\cite{Motoba1985PTPS}.

The calculated $B(E1)$ values are listed in Tables \ref{C-hyperon} for C isotopes and \ref{B-hyperon}  for B isotopes, respectively.  The values are larger in light isotopes than those in heavier nuclei  because of the effective charge in  Eq. \eqref{eq:Lambda-BE1-effelam}.  
  The $B(E1:1p_{3/2}\rightarrow 1s_{1/2})$=0.1036 $e^2$fm$^2$ of hyperon configurations in $^{13}_{\Lambda}$C corresponds to 0.29$B_W(E1)$, where $B_W(E1)$ is the Weisskopf unit (single-particle unit) of electric dipole transition in $A=13$ nucleus.   The decay half-life $t_{1/2}$ is estimated as
\be
t_{1/2}=\frac{\ln 2}{T(E1)}=2.99\times 10^{-18} ~{\rm sec}, 
\ee
where $T$ is the decay rate,
\be
T(E1)=1.59\times 10^{15}(E_x)^3B(E1)=2.31\times 10^{17} {\rm sec^{-1}}.
\ee
The $T(E1)$ is evaluated to be 1.51$\times 10^{17}$ sec$^{-1}$ for the transition $(\Lambda( 1p_{3/2})\rightarrow \Lambda( 1 s_{1/2}))$ in $^{13}_{\Lambda}$B and the half-life is estimated as $t_{1/2}$= 4.60$\times 10^{-18}$ sec.

 In halo nuclei without $\Lambda$ degree of  freedom, the largest $B(E1)$ transition between discrete states is observed in $2s_{1/2}\rightarrow 
1p_{1/2}$ transition in $^{11}$Be~\cite{Nakamura1}; $B(E1;2s_{1/2}\rightarrow 1p_{1/2})=0.099\pm0.010 e^2$fm$^2 =0.31\pm0.03 B_W(E1)$, 
which is almost the same strength as $B(E1:\Lambda(1 p_{3/2})\rightarrow \Lambda(1 s_{1/2}))$ of hyperon configurations in $^{13}_{\Lambda}$C.  
Notice these $B(E1)$ in halo nuclei (hypernuclei) are 2-3 order of magnitude larger than normal $B(E1)$, which is less than 10$^{-3}e^2$fm$^2$.  
The $B(E1)$ strength of  halo  nuclei was studied also by the Coulomb breakup reactions, which measure the excitation from the halo state to the continuum. 
In these reactions, the $B(E1)$ value was found ${B(E1: {\rm exp})=1.05\pm0.06 e^2}$fm$^2$ in $^{11}$Be \cite{Fukuda2004} and
${B(E1:{\rm exp})=0.71\pm 0.07 e^2}$fm$^2$ in $^{19}$C \cite{Nakamura2}. 
Systematic measurements  of electromagnetic transitions in $\Lambda(1p)$ states may give us a peculiar nuclear structure information including the characteristic 
 features of hyperon halo wave functions.  

Here we should mention that, the present Skyrme Hartree-Fock model is not suitable for the very weakly bound states. Instead, the Hartree-Fock-Bogoliubov model with pairing correlation and continuum effects~\cite{Dobaczewski1984NPA, Meng2006PPNP} will be more reliable for these states.  However in the present investigation, we apply the simple Hartree-Fock model as the first step, since the single-particle wave function is straightforward to calculate the transition probability as shown in Eq.~(\ref{eq:Lambda-sp-BE1}).  The next step to include the pairing and continuum effects is in progress.

\section{Summary and future perspectives}

In this work, we calculated the $\Lambda$ single-particle states systematically in the C and B isotopes using the HF approach with the Skyrme-type $\Lambda N$ interaction derived from the $G-$matrix calculation of the one-boson-exchange potential.  We tuned the strength of $\Lambda N$ spin-orbit interaction by fitting to the observed spin-orbit splitting data of $1/2^- - 3/2^-$ states in $^{13}_{\Lambda}$C.   The $\Lambda$ binding energies thus obtained agree with the available experiment data quite well for the C and B  hypernuclei.  In the light hypernuclei $^{12-14}_{~~~~\Lambda}$C and $^{12-14}_{~~~~\Lambda}$B, we found very weakly bound excited $1p$ orbits for $\Lambda$ hyperon, which could have much extended density and large rms radii compared with the ground $1s$ state.   
Furthermore, we calculated $B(E1)$ values.  This halo structure may provide the enhanced $E1$ transition from the excited $1p$ states to the ground $1s$ state, which is a challenging open problem for the future experiment.  On the other hand, with more neutrons, the $\Lambda$ levels become more deeply bound, so that the hyperon halo structure disappears.


\begin{acknowledgments}
This work was supported by  JSPS KAKENHI  Grant Numbers  JP19K03858,  JP18H05407,  and China Scholarship Council (Grant No. 201906255002).  
\end{acknowledgments}

\begin{appendix}
\end{appendix}



\clearpage

\end{document}